\documentclass[letterpaper,12pt]{article}
\usepackage{color,amssymb,enumerate,float}
\usepackage{amsmath}

\usepackage{ifpdf}
\ifpdf
	\usepackage[pdftex]{graphicx}
	\usepackage[pdftex,unicode,implicit]{hyperref}

	\hypersetup{
  	pdftitle     = {}, 
  	pdfkeywords  = {},
  	pdfauthor    = {},
  	pdfcreator   = {pdf\LaTeXe\ with package \flqq hyperref\frqq},
  	pdfproducer  = {pdf\LaTeXe\ with package \flqq hyperref\frqq},
  	pdfpagemode  = UseNone,  
  	pdffitwindow = true,  
  	unicode      = true,
  	plainpages   = true,
  	colorlinks   = true,  
  	citecolor    = black,  
  	urlcolor     = blue, 
  	linkcolor    = black
	}

\else

  \usepackage[dvips]{graphicx}

\fi 

\makeatletter
\@addtoreset{equation}{section}
\makeatother

\begin{document}
	
\thispagestyle{empty}

\begin{center}
{\bf \LARGE Quantization of the anisotropic conformal Ho\v{r}ava theory}
\vspace*{15mm}

{\large Jorge Bellor\'{\i}n}$^{1,a}$,
{\large Claudio B\'orquez$^{1,b}$}
{\large and Byron Droguett}$^{2,c}$
\vspace{3ex}

$^1${\it Department of Physics, Universidad de Antofagasta, 1240000 Antofagasta, Chile.}

$^2${\it Departamento de Ciencias B\'asicas, Facultad de Ciencias, Universidad Santo Tom\'as, Sede Arica 1000000, Chile.}
\vspace{3ex}

$^a${\tt jorge.bellorin@uantof.cl,} \hspace{1em}
$^b${\tt cl.borquezg@gmail.com,} \hspace{1em}
$^c${\tt byrondroguett@santotomas.cl}

\vspace*{15mm}
{\bf Abstract}
\begin{quotation}{\small} 
  We perform the Batalin-Fradkin-Vilkovisky quantization of the anisotropic conformal Ho\v{r}ava theory in $d$ spatial dimensions. We introduce a model with a conformal potential suitable for any dimension.	We define an anisotropic and local gauge-fixing condition that accounts for the spatial diffeomorphisms and the anisotropic Weyl transformations. We show that the BRST transformations can be expressed mainly in terms of a spatial diffeomorphism along a ghost field plus a conformal transformation with another ghost field as argument. We study the quantum Lagrangian in the $d=2$ case, obtaining that all propagators are regular, except for the fields associated with the measure of the second-class constraints. This behavior is qualitatively equal to the nonconformal case.
\end{quotation}
\end{center}

\thispagestyle{empty}

\newpage
\section{Introduction}
The Ho\v{r}ava theory \cite{Horava:2009uw} provides a scenario for studying quantum gravity with anisotropic symmetries between time and space, resulting in the breaking of the local Lorentz symmetry. The anisotropy is introduced with the aim of having a renormalizable theory of quantum gravity, which is a fundamental open question in theoretical physics, and simultaneously with quantum unitarity. The formulation of the theory starts with the assumption of the existence of a foliation of spatial slices along a given direction of time, such that the foliation has an absolute physical meaning. The gauge symmetry is restricted to the diffeomorphisms that preserve the given foliation. On the basis of this anisotropic gauge symmetry one may introduce terms in the Lagrangian with higher order in spatial derivatives, while keeping the time derivatives at order two. In this way, the resulting theory is expected to be renormalizable and unitary. Indeed, the renormalization of the projectable version has been proven \cite{Barvinsky:2015kil}.

There is a formulation of the Ho\v{r}ava theory with a kind of Weyl symmetry that is also anisotropic. This conformal version was presented in the original formulation of the theory in Refs.~\cite{Horava:2009uw} and \cite{Horava:2008ih}. Thus, this formulation gives an interesting model of study of the Weyl symmetry that is qualitatively different to the case of the Weyl symmetry in general relativity. As a consequence of its anisotropic nature, the Lagrangian of the Ho\v{r}ava theory is divided between its kinetic and potential parts. The anisotropic Weyl symmetry in the kinetic part arises at a critical value of the corresponding coupling constant. For the potential, one must take models that are conformal. Since the Weyl transformations in the potential concern only spatial derivatives, the conformal potentials are rather different to the case of conformal general relativity. A well-known model on a foliation of three spatial dimensions is the square of the Cotton tensor, introduced in \cite{Horava:2009uw}.

In the context of the general Ho\v{r}ava theory without extra gauge symmetries, there have been advances on its quantization. The nonprojectable case is the version with the richer dynamics and it is closer to general relativity at low energies. An essential extension of the Lagrangian of nonprojectable case was presented in Ref.~\cite{Blas:2009qj}. We remark that the anisotropic Weyl symmetry is defined exclusively within the nonprojectable case. The nonprojectable case has second-class constraints, a feature that demands a careful treatment of the quantization. A central step to achieve this is the incorporation of the quantum measure associated with these constraints \cite{Senjanovic:1976br,Bellorin:2019gsc}. In this sense, a general formalism is the Batalin-Fradkin-Vilkovisky (BFV) quantization \cite{Fradkin:1975cq,Batalin:1977pb,Fradkin:1977xi}. In previous papers we have undertaken the quantization of the general Ho\v{r}ava theory under this formalism \cite{Bellorin:2021tkk,Bellorin:2021udn,Bellorin:2022qeu,Bellorin:2022efu}. The BFV formalism is suitable to incorporate the second-class constraints together with the appropriate gauge-fixing condition for renormalization. This gauge-fixing condition is noncanonical in the sense of the classical Hamiltonian formalism \cite{Barvinsky:2015kil,Bellorin:2021udn}. Therefore, the BFV formalism is a promissory scenario for the consistent quantization of the (nonprojectable) Ho\v{r}ava theory, in which the problem of its renormalization can be addressed.

Following these advances, an interesting question is whether the anisotropic Weyl symmetry can be incorporated in the consistent quantization. In this paper we develop this study, by analyzing the BFV quantization of the conformal Ho\v{r}ava theory in arbitrary $d$ spatial dimensions. Since the conformal theory belongs to the nonprojectable version, it has second-class constraints. Thus, the quantization program requires the handling of the second-class constraints and a gauge fixing-condition that takes into account the extra gauge symmetry of the anisotropic Weyl transformations. Our approach consists of making the required extension of the phase space for the BFV quantization and studying the formulation of the path integral. The presence of the extra gauge symmetry requires a definition of the gauge-fixing condition different to the general nonconformal case.

The anisotropic conformal Ho\v{r}ava theory has been studied in several ways. A related version is when the coupling constant of the kinetic term takes its critical value (the theory is at the critical point); hence the kinetic term is conformal, but the potential is not conformal (the name kinetic-conformal Ho\v{r}ava theory has been used for this version). Among the studies on these theories, an analysis of the degrees of freedom was performed in Ref.~\cite{Park:2011pao}, taking the theory at the critical point and a combination of conformal and nonconformal terms in the potential. Two of us developed in Ref.~\cite{Bellorin:2018blt} a comparative analysis between the classical dynamics of the conformal theory and the theory at the critical point. It is interesting that both theories have many features in common. Mainly, in both cases the extra mode typically associated with the Ho\v{r}ava theory is eliminated, hence both version share the same number of degrees of freedom with general relativity. In the case of the exact conformal theory, this is a consequence of the extra gauge symmetry of anisotropic Weyl transformations. In the theory at the critical point, there is no such a symmetry in an exact way, but there arise two extra second-class constraints, one of them being the generator of the Weyl transformations, that eliminate the extra mode \cite{Bellorin:2013zbp}. Interestingly, the critical point for the anisotropic Weyl symmetry has been found to be an UV fixed point in the renormalization flow of the projectable Ho\v{r}ava theory in $2+1$ dimensions \cite{Barvinsky:2017kob}.

It is widely known that in relativistic theory with conformal symmetry the Weyl anomaly plays a prominent role. The anomaly is also present in the anisotropic case: in Ref.~\cite{Adam:2009gq} the Weyl anomaly was found for the case of an anisotropic theory of a scalar field (a Lifshitz scalar), living on a Horava-gravity background. Hence, one expects that the anisotropic Weyl symmetry of the conformal Horava theory is anomalous. We remark that in this paper we study the quantization of the purely gravitational theory. Other studies of quantization of the gravitational theory with coupling to matter fields and the Weyl anomaly can be developed as well. Indeed, the anisotropic conformal Ho\v{r}ava gravity has arisen in holography with asymptotic Lifshitz scaling, where it seems to be a natural scenario for the holographic duality. Holographic renormalization with asymptotic Lifshitz scaling was studied in Ref.~\cite{Griffin:2011xs}, finding that the holographic counterterms are associated to the Ho\v{r}ava gravity. In particular, in \cite{Griffin:2011xs} it was found that the anisotropic Weyl anomaly of the conformal field theory at the boundary is related precisely to the action of the conformal Ho\v{r}ava gravity. The general duality between Ho\v{r}ava gravity in the bulk and field theories with Lifshitz scaling at the boundary has been proposed in \cite{Griffin:2012qx}.

This paper is organized as follows. In section 2 we summarize the classical formulation of the anisotropic conformal Ho\v{r}ava gravity, including the Hamiltonian formulation, with a specific conformal potential. In section 3 we develop the BFV quantization of the same model. In section 4 we study in detail the BRST symmetry resulting from the quantization. In section 5 we present the quantum Lagrangian obtained by integration on the canonical momenta, with the propagators of the fields, in the case of the $2+1$ conformal theory. We finally present some conclusions. In the appendix we develop the quantization of model with broken conformal symmetry in $3+1$ dimensions.


\section{Anisotropic Weyl symmetry}
The Ho\v{r}ava theory is based on the existence of a foliation of $d$-dimensional spatial slices along a definite direction of time. The foliation has an absolute physical meaning. In the Lagrangian formulation, the fields representing the gravitational interaction are the Arnowitt-Deser-Misner variables $g_{ij}$, $N$ and $N^i$. The basic underlying gauge symmetry on these variables is given by the diffeomorphisms that preserve the foliation (FDiff).

The classical action of the general theory in Lagrangian form over a foliation of $d$-dimensional spatial slices is \cite{Horava:2009uw,Blas:2009qj}
\begin{equation}
 S = 
 \int dt d^dx \sqrt{g} N \left ( 
  K_{ij} K^{ij} - \lambda K^2 - \mathcal{V} \right) \,,
\end{equation}
where $K_{ij}$ is the extrinsic curvature tensor of the spatial slices,
\begin{equation}
 K_{ij} = \frac{1}{2N} \left ( 
 \dot{g}_{ij} - 2 \nabla_{(i} N_{j)} \right) \,.
\end{equation}
The dot denotes derivative with respect the time, as usual. $\lambda$ is the coupling constant of the kinetic term. Its value plays a central role in the definition of the conformal theory. $\mathcal{V}$ denotes the potential of the theory. For the case of the general theory $\mathcal{V}$ contains all the nonequivalent terms compatible with the FDiff symmetry that depend on the spatial metric $g_{ij}$ and the FDiff-covariant vector $a_i = \partial_i \ln N$. Spatial derivatives are included in $\mathcal{V}$ up to the order $2z$, where $z=d$ as required by power-counting renormalization.

The definition of the (anisotropic) conformal theory \cite{Horava:2009uw} starts with the observation that if $\lambda$ takes its critical value
\begin{equation}
 \lambda = \frac{1}{d} \,,
\end{equation}
then anisotropic Weyl scalings, defined by 
\begin{equation}
\tilde{g}_{ij} = \Theta^2 g_{ij} \,,
\hspace{2em}
\tilde{N} = \Theta^{d} N \,,
\hspace{2em}
\tilde{N}_i = \Theta^2 N_i \,,
\label{introweyl}
\end{equation}
where  $\Theta = \Theta(t,\vec{x})$, render the kinetic term of the Lagrangian conformally covariant,
\begin{equation}
\tilde{K}_{ij} \tilde{K} ^{ij}
- \frac{1}{d}\tilde{K}^2
= \Theta^{-2d}\left(K_{ij} K ^{ij}-\frac{1}{d}K^2\right) \,.
\end{equation}
Since the volume element transforms as $\sqrt{\tilde{g}} \tilde{N} = \Theta^{2d} \sqrt{g} N$, the kinetic sector of the Lagrangian is left invariant by (\ref{introweyl}). The conformal transformations in (\ref{introweyl}) are anisotropic in the sense that the weight assigned to the scaling of the lapse function differs from the ones of the spatial metric and the shift vector. Note that this transformation applies only to the nonprojectable Ho\v{r}ava theory, since $N$ depends both in time and space. Hence the nonprojectable case is our subject of study in this paper.

To have a complete conformal theory, the potential $\mathcal{V}$ must also be conformal. A well--known case in $d=3$ is the $(\text{Cotton--tensor})^2$ potential \cite{Horava:2009uw}. Here we present another conformal potential, whose form can be adapted to any spatial dimension $d$. An important feature of this model is its dependence on the lapse function $N$, which is quite relevant for the consistency of the dynamics (see Ref.~\cite{Bellorin:2018blt} for the $d=3$ case). According to (\ref{introweyl}), the conformal transformation of the FDiff-covariant vector $a_i$ is given by
\begin{equation}
    \tilde{a}_i = a_i + d \partial_i \ln(\Theta) \,.
\end{equation}
This has the form of a conformal gauge connection. We propose the following  tensor
\begin{eqnarray}
    \chi_{ij} = R_{ij} + \alpha_1 a_i a_j + \alpha_2 \nabla_{(i}a_{j)},
\end{eqnarray}
for any spatial dimension $d$. The constants $\alpha_1 , \alpha_2$ are parameters that can be adjusted to get the conformal invariance. This tensor transforms  under anisotropic conformal scalings as
\begin{eqnarray}
    \tilde{\chi}_{ij} &=&
    \chi_{ij}
    + (2\alpha_1 d-2\alpha_2)\Theta^{-1}a_{(i}\nabla_{j)}\Theta
    + \alpha_2 g_{ij}\Theta^{-1}a^k\nabla_k\Theta
    + (2-d+\alpha_2 d)\Theta^{-1}\nabla_i\nabla_j\Theta
    \nonumber\\
    &&
    - g_{ij}\Theta^{-1}\nabla^2\Theta
    + (2d-4+\alpha_1 d^2-3d\alpha_2)\Theta^{-2}\nabla_i\Theta\nabla_j\Theta
    \nonumber\\
    &&
    + (3-d+d\alpha_2)g_{ij}\Theta^{-2}\nabla^k\Theta\nabla_k\Theta \,.
\end{eqnarray}
Its trace $\chi = g^{ij} \chi_{ij}$ transforms as
\begin{eqnarray}
\tilde{\chi} &=&
    \Theta^{-2}\chi
    + \left(2\alpha_1 d+\alpha_2(d-2)\right)\Theta^{-3}a^k\nabla_k\Theta
    + \left(2(1-d)+\alpha_2 d\right)\Theta^{-3}\nabla^2\Theta
    \nonumber\\
    &&
    + \left(\alpha_1 d^2+\alpha_2 d(d-3)-(d-1)(d-4)\right)\Theta^{-4}\nabla^k\Theta\nabla_k\Theta \,.
    \label{traza}
\end{eqnarray}
The first term on the right-hand side suggests that the trace can be a conformal object of conformal weight $-1$,  $\tilde{\chi}=\Theta^{-2}\chi$. To achieve this, the coefficients of the rest of terms must vanish. It turns out that this is satisfied with the setting
\begin{equation}
    \alpha_1 = 
    \frac{(2-d)(d-1)}{d^2} \,,
    \quad 
    \alpha_2 = 
    \frac{ 2 (d-1) }{ d } \,.
    \label{alpha1alpha2}
\end{equation}
Once $\alpha_1$ and $\alpha_2$ are set to these values, we may define the anisotropic conformal potential as
\begin{equation}
 \mathcal{V}=\varrho\chi^d \,,
 \label{potential}
\end{equation}
where $\varrho$ is an arbitrary coupling constant. This potential transforms as $\tilde{\mathcal{V}}=\Theta^{-2d}\mathcal{V}$. Some cases of $\chi$ are
\begin{eqnarray}
&& 
d=2 \,: \quad \chi =
R + \nabla_i a^i \,,
\label{potentiald2}
\\ && 
d=3 \,: \quad \chi =
R - \frac{2}{9} a_i a^i + \frac{4}{3} \nabla_i a^i \,.
\end{eqnarray}
The action of the anisotropic conformal Ho\v{r}ava theory with this new conformal potential is given by
\begin{equation}
    S = \int\,dt\,d^dx \sqrt{g} N
    \Big(G^{ijkl}K_{ij}K_{kl} - \varrho\mathcal{\chi}^d \Big) \,,
\end{equation}
where $G^{ijkl} = \frac{1}{2} ( g^{ik} g^{jl} + g^{il} g^{jk} ) - \frac{1}{d} g^{ij} g^{kl}$. 

The Hamiltonian formulation of the general nonprojectable Ho\v{r}ava theory without extra symmetries was presented in Refs.~\cite{Kluson:2010nf,Donnelly:2011df,Bellorin:2011ff}, showing the consistence of the classical dynamics. The canonical momentum conjugated to the metric is
\begin{eqnarray}
\pi^{ij} = \frac{\delta\mathcal{L}}{\dot{g}_{ij}} = \sqrt{g}G^{ijkl}K_{kl}\,.
\end{eqnarray}
The hypermatrix $G^{ijkl}$ is not invertible when $\lambda=1/d$, since, in general,
\begin{eqnarray}
G^{ijkl}g_{ij} = \left(1-d\lambda\right)g^{kl} = 0 \,.
\end{eqnarray}
Therefore, for the formulation of the conformal Ho\v{r}ava theory we must move $\lambda$ to its critical point $\lambda = 1/d$; hence we obtain the primary constraint
\begin{equation}
 \pi = 0 \,,
 \quad
 \pi \equiv g_{ij} \pi^{ij} \,.
\end{equation}
This is the generator of the conformal transformations on the fields $( g_{ij} , \pi^{ij})$; hence it is expected to have it as part of the constraints. The momentum conjugated to the lapse function is  
\begin{eqnarray}
P_N = \frac{\delta\mathcal{L}}{\delta \dot{N}} = 0
\end{eqnarray}
which is another primary constraint. The conservation of the $P_N=0$ leads to the  secondary constraint
\begin{equation}
 \theta_1 = 
 \frac{N}{\sqrt{g}}\pi^{ij}\pi_{ij}
 + \varrho \sqrt{g} N \chi^d
 + 2\varrho(d-1) \sqrt{g} \left(\nabla^2(N\chi^{d-1})
 + \frac{1}{d}(d-2)\nabla_k(a^k N\chi^{d-1} )\right) \,.
 \label{prehamiltonian}
\end{equation}
The last terms in Eq.~(\ref{prehamiltonian}) can be written in terms of a conformal covariant derivative. The scalar factor $N\chi^{d-1}$ is conformal
\begin{equation}
    \tilde{N}\tilde{\chi}^{d-1} = \Theta^{2-d}N\chi^{d-1} \,.
\end{equation}
Therefore, the expression
\begin{equation}
    D_k\left(N\chi^{d-1}\right) \equiv 
    \partial_k\left(N\chi^{d-1}\right) 
    + \frac{d-2}{d} a_k N\chi^{d-1} \,,
\end{equation}
transforms as
\begin{equation}
    \tilde{D}_k\left(\tilde{N}\tilde{\chi}^{d-1}\right) = \Theta^{2-d}D_k\left(N\chi^{d-1}\right) \,.
\end{equation}
Hence the last term of Eq.~(\ref{prehamiltonian}) is equal to $ 2\varrho(d-1) \sqrt{g} \nabla_i D^i \left(N\chi^{d-1}\right)$.
These results are useful to check the conformal invariance of $\theta_1$ (see Ref.~\cite{Bellorin:2018blt} for the $d=3$ case).
As a consequence of the conformal symmetry, the preservation of the constraint $\pi$ does not yield new constraints. The $d+1$-dimensional conformal nonprojectable theory has the following constraints
\begin{eqnarray}
    &&
    \mathcal{H}_i =
    - 2g_{ij}\nabla_k \pi^{jk}  = 0\,,
    \label{vinmomen}
    \\ &&
    \pi = 0 \,, 
    \label{pi0}
    \\ &&
    \theta_{1} = 
    \frac{N}{\sqrt{g}}\pi^{ij}\pi_{ij}
    + \varrho\sqrt{g}N\chi^d
    + 2\varrho(d-1)\sqrt{g} \nabla_k D^k\left(N\chi^{d-1}\right) 
    = 0 \,,
    \label{theta1}
    \\ &&
    \theta_{2} = P_{N} = 0 \,.
    \label{theta2}
\end{eqnarray}
$\theta_1$ and $\theta_2$ are second-class constraints. The Hamiltonian with all constraints incorporated is given by
\begin{eqnarray}
    H &=&   \int\,d^dx\left[
    \frac{N}{\sqrt{g}}\pi^{ij}\pi_{ij}
    + \varrho\sqrt{g}N\left(R + \frac{1}{d^2}(2-d)(d-1)a_i a^i + \frac{2}{d}(d-1)\nabla_ka^k\right)^d
    \right.
    \nonumber
    \\
    &&
    + N^i \mathcal{H}_i
    + \mu\pi
    + \mathcal{A} \theta_{1}
    + \mathcal{B} \theta_{2} \Big] \,,
 \label{Hamiltonianclassical}
\end{eqnarray}
where $N^i$, $\mu$, $\mathcal{A}$, and $\mathcal{B}$ are Lagrange multipliers. Constraint $\theta_{1}$ contains a total spatial derivative. Once we integrate it, this term vanishes with appropriated boundary conditions. Therefore, we can prove that the primary Hamiltonian is equivalent to the integral of a second-class constraint,
\begin{equation}
	H_0 =
	\int d^dx\, \theta_1 \,.
	\label{Hamiltonianconstraint}
\end{equation}

The canonical formulation requires the extension of the anisotropic conformal transformations in the way
\begin{equation}
\begin{array}{ll}
\tilde{g}_{ij} = \Theta^2g_{ij}\,,
\hspace*{4em}
&
\tilde{\pi}^{ij} =\Theta^{-2}\pi^{ij}
\,,
\\[1ex]
\tilde{N} = \Theta^dN\,,
&
\tilde{P}_N =\Theta^{-d}\theta_{2}  \,,
\\[1ex]
\tilde{N}_{k} = \Theta ^2N_k\,.
\\[1ex] 
\end{array}
\end{equation}
The invariance of the canonical action under these transformations requires specific transformations of the rest of Lagrange multipliers. Under a conformal transformation the canonical action becomes
\begin{eqnarray}
\tilde{S} &=&
\int\,dt\,d^{d}x 
\left[ 
P_N \dot{N}
+ \pi^{ij}\dot{g_{ij}}
- \frac{N}{\sqrt{g}}\pi^{ij}\pi_{ij}
- \varrho\sqrt{g}N\chi^d 
- N^i \mathcal{H}_i
\right.
\nonumber\\
&+&
\Big( d \Theta^{-1}\dot{\Theta}N
- \tilde{\mathcal{B}} \Theta^{-d}\Big) \theta_{2}
+ \pi\left(2\Theta^{-1}\dot{\Theta}-2\Theta^{-1}N_k\partial^k\Theta-\tilde{\mu}\right)
-\tilde{\mathcal{A}}\theta_{1}\Big] \,.
\end{eqnarray}
Therefore, we require the Lagrange multipliers to transform as 
\begin{eqnarray}
\tilde{\mu} &=&
\mu+2\Theta^{-1}\dot{\Theta}
-2 \Theta^{-1} N^i \partial_i \Theta \,,
\\
\tilde{\mathcal{B}} &=&
\Theta^{d} \mathcal{B} +dN\dot{\Theta}\Theta^{d-1}  \,,
\\
\tilde{\mathcal{A}} &=& \mathcal{A}\,.
\end{eqnarray}

The phase space of the classical theory is spanned by the variables $(g_{ij},\pi^{ij})$ and $(N,P_N)$, which contain $d(d+1) + 2$ degrees of freedom. There are two first-class constraints, $\mathcal{H}_i$ and $\pi$, with two gauge symmetries associated, and two second-class constraints, $\theta_1$ and $\theta_2$. Hence, $2(d+1)$ degrees must be eliminated as unphysical from the phase space. The resulting canonical theory has a number of $\frac{1}{2} d (d-1) - 1$ physical modes. This number coincides with general relativity; the so called extra mode of the Ho\v{r}ava theory is eliminated by the gauge symmetry of anisotropic conformal transformations. In $d=2$ the conformal theory propagates no physical modes and in $d=3$ it propagates two physical modes.


\section{BFV quantization}
We utilize the definition of Dirac brackets,
\begin{equation}
\left\lbrace F,R\right\rbrace_{\text{D}} = 
\left\lbrace F,R\right\rbrace
- \left\lbrace F,\theta_{A}\right\rbrace\mathbb{M}^{-1}_{AB}\left\lbrace\theta_{B},R\right\rbrace\,,
\;\;\;
\mathbb{M}_{AB} = \left\lbrace\theta_{A},\theta_{B}\right\rbrace\,.
\end{equation}
The constraints that satisfy an involutive algebra with respect to the Dirac brackets are $\mathcal{H}_i$ and $\pi$. Their brackets are given by 
\begin{eqnarray}
\{\pi(x),\pi(y)\}_{\text{D}} &=& 
0 \,,
\label{pipi}
\\
\{ \pi(x), \mathcal{H}_i (y) \}_{\text{D}} &=& 
\frac{ \partial \delta(x-y) }{ \partial x^i } \pi(y) \,,
\label{Hipi}
\\
\{\mathcal{H}_i(x),\mathcal{H}_j(y)\}_{\text{D}} &=&
\frac{\partial \delta(x-y) }{\partial x^i}\mathcal{H}_j(x)
- \frac{\partial \delta(x-y) }{\partial y^j}\mathcal{H}_i(y)\,.
\label{HiHj}
\end{eqnarray}
The coefficients of this involutive algebra enter in the definition of the BRST charge. 
 
The BFV extension of the phase space is done as follows: the Lagrange multipliers of the involutive constraints $\mathcal{H}_i$ and $\pi$ are promoted to canonical variables together with their conjugate momenta. They define the new canonical pairs $(N^{i},\pi_{i})$ and $(\mu,Q)$. The BFV ghosts associated with spatial diffeomorphisms are $( C^{i}, \bar{\mathcal{P}}_i )$ and $( \mathcal{P}^i , \bar{C}_i )$. The BFV ghosts associated with the conformal symmetry are $( C, \bar{\mathcal{P}})$ and $( \mathcal{P}, \bar{C} )$. A useful collective notation for all these ghosts is $\eta^{a}=( C^i,\mathcal{P}^i,C,\mathcal{P})$, $\mathcal{P}_{a}=(\bar{\mathcal{P}}_{i}, \bar{C}_i,\bar{\mathcal{P}}, \bar{C})$. Thus, the full phase space is given by the canonical pairs $(g_{ij},\pi^{ij})$, $(N,P_{N})$, $(N_{i},\pi^{i})$, $(\mu,Q)$, and $(\eta_{a},\mathcal{P}^{a})$.

The BFV path integral of the nonprojectable Ho\v rava theory is given by
\begin{equation}
	Z = \int\mathcal{D}V e^{iS}\,,
\end{equation}
where the measure and the canonical action are, respectively,
\begin{eqnarray}
	\mathcal{D}V &=&
	\mathcal{D}g_{ij} \mathcal{D} \pi^{ij} \mathcal{D}N \mathcal{D}P_{N} \mathcal{D}N^{k} \mathcal{D}\pi_{k}\mathcal{D} \mu \mathcal{D}Q \mathcal{D}\eta^{a} \mathcal{D} \mathcal{P}_{a}
	\times \delta(\theta_{1})\delta(\theta_{2}) \sqrt{\det\mathbb{M}} \,,
	\label{measureoriginal}
	\\
	S &=& \int dtd^{d}x
	\left(
	\pi^{ij}\dot{g}_{ij}
	+ P_{N}\dot{N}
	+ \pi_{k}\dot{N}^{k}
	+ Q\dot{\mu}
	+ \mathcal{P}_{a}\dot{\eta}^{a}
	- \mathcal{H}_{\Psi}
	\right)\,.
	\label{principalaction}
\end{eqnarray}
The quantum gauge-fixed Hamiltonian is
\begin{equation}
	\mathcal{H}_{\Psi} = \mathcal{H}_0 + \left\lbrace\Psi,\Omega\right\rbrace_{\text{D}} \,,
\end{equation}
where $\Psi$ is the fermionic function chosen to fix the gauge symmetry and $\Omega$ is the BRST charge. In the case of the conformal Ho\v{r}ava theory, $\Omega$ admits an expansion up to linear order on the BFV ghosts $\mathcal{P}_a$ (the theory is of order one \cite{Fradkin:1977xi}). Its general definition, using symbolic notation, is
\begin{equation}
\Omega =
G_{a}\eta^{a}
- \frac{1}{2}U^{c}_{ab}\eta^{a}\eta^{b}\mathcal{P}_{c}
\,,
\end{equation}
where $G_a$ denotes the functions $G_{a}=(\mathcal{H}_{i},\pi_{i},\pi,Q)$, and $U_{ab}^c$ denotes the coefficients of the algebra between the constraints $\mathcal{H}_i$ and $\pi$, given in Eqs.~(\ref{pipi})--(\ref{HiHj}). Explicitly, the BRST charge results in
\begin{equation}
	\Omega =
    \int d^dx \left(\mathcal{H}_{k}C^{k}
	+ \pi_{k}\mathcal{P}^k
	+ \pi C+Q\mathcal{P}-C^k\partial_kC^l\bar{\mathcal{P}}_l-\partial_k(C^k\bar{\mathcal{P}})C\right)
	\,.
\end{equation}
$\Omega$ must satisfy the consistency conditions required in the BFV formalism, which are
\begin{equation}
 \{ \Omega \,, \Omega \}_{\text{D}} = 0 \,,
 \qquad
 \{ \mathcal{H}_0 \,, \Omega \}_{\text{D}} = 0 \,.
\end{equation}
The first condition can be checked by direct computations, whereas the second one holds due to that $\mathcal{H}_0$, taken as a differentiable functional, it is equivalent to a second-class constraint, hence its Dirac brackets are always zero.

To fix the gauge we use perturbative variables, denoted by
\begin{equation}
 g_{ij} = \delta_{ij} + h_{ij} \,,
 \quad
 \pi^{ij} = p_{ij} \,,
 \quad
 N = 1 + n \,,
 \quad
 N^i = n^i \,.
\end{equation}
On the rest of field variables we keep the original notation, with perturbative meaning. In the original BFV formulation a generic form of the gauge-fixing condition (for an arbitrary gauge symmetry) was introduced. Its functional form was aimed to apply it to relativistic systems with covariant (and noncanonical) gauge-fixing conditions. This leads to a generic form of the fermionic function $\Psi$, which we adopt for the conformal Ho\v{r}ava theory. Function $\Psi$ is
\begin{equation}
\Psi = 
  \bar{\mathcal{P}}_i n^i
+ \bar{C}_i \chi_1^i
+ \bar{\mathcal{P}}\mu
+ \bar{C}\chi_2 \,,
\end{equation}
where $\chi_1^i$ and $\chi_2$ are functionals of the canonical fields to be chosen freely. It seems natural to take $\chi_1^i$ to fix the symmetry of spatial diffeomorphisms, whereas $\chi_2$ does the same for the conformal transformations (although these symmetries are not independent). In previous analysis \cite{Bellorin:2021udn}, we have introduced a gauge-fixing condition for the symmetry of the spatial diffeomorphisms that depends on the momentum $\pi_i$. The idea here is to follow the same strategy; hence we use $\pi_i$ in the sector of the spatial diffeomorphisms and the momentum $Q$ in the sector of the conformal transformation. Thus, we set
\begin{eqnarray}
&&
	\chi^{i}_1 = 
	\sigma_{1}\mathfrak{D}^{ij}\pi_{j}
	+ \Gamma^{i}_1 \,,
\label{chi1}
\\ && 
	\chi_2 = 
	\sigma_2\Delta^d Q
	+ \Gamma_2 \,,
\label{gauge2}
\end{eqnarray}
where
\begin{eqnarray}
&&
\Delta = \partial_{kk} \,,
\quad
\mathfrak{D}^{ij} = 
\delta_{ij}\Delta^{d-1}
+ \kappa\Delta^{d-2}\partial_{ij} \,,
\\ &&
 \Gamma^i_1 = 
   2c_1 \Delta^{d-1} \partial_j h_{ij}
 + 2c_2 \Delta^{d-1} \partial_i h
 + c_3 \Delta^{d-2} \partial_{ijk} h_{jk} \,,
\\ &&
 \Gamma_2 =
   c_4 \Delta^{d} h + c_5 \Delta^{d-1} \partial_{ij} h_{ij}\,.
\end{eqnarray}
$\mathfrak{D}^{ij}$ is the appropriate operator to connect with the nonlocal quantum Lagrangian \cite{Barvinsky:2015kil,Bellorin:2021udn}. We have chosen $\Gamma_1^i$ and $\Gamma_2$ in the more general form allowed by the  anisotropic scaling of the fields. At this stage $\sigma_1$, $\sigma_2$, $\kappa$, and $c_{1,\ldots,5}$ are arbitrary constants. Below we see that the setting \cite{Barvinsky:2015kil,Bellorin:2021udn}
\begin{eqnarray}
c_{1}=-\sigma_{1}\,, \qquad c_{2}=0\,, \qquad c_{3}=-2\kappa\sigma_{1}\,, \qquad c_{4}=\sigma_{2}\,, \qquad c_{5}=0 \,,
\end{eqnarray}
leads to an important simplification in the Lagrangian: odd derivatives in time and space are eliminated. We use in advance this fact, so the final form of the gauge-fixing factors is
\begin{eqnarray}
 &&
 \Gamma^i_1 = 
 - 2 \sigma_1 \left(
 \Delta^{d-1} \partial_j h_{ij}
 + \kappa \Delta^{d-2} \partial_{ijk} h_{jk} 
 \right) \,,
  \label{Gamma1}
 \\ &&
 \Gamma_2 =
 \sigma_2 \Delta^{d} h \,.
\end{eqnarray}
Notice that with this setting $\Gamma_1^i$ gets contributions of the longitudinal sector of $h_{ij}$ exclusively, which is related to symmetry of spatial diffeomorphisms. Moreover, $\Gamma_2$ only involves the trace $h$, which is naturally associated to the conformal symmetry, but it does not get contributions from the vectorial longitudinal sector. The gauge-fixed Hamiltonian take the form
\begin{eqnarray}
{\mathcal{H}}_{\Psi}
&=&
\mathcal{H}_{0} + \mathcal{H}_{k} n^{k}
+ \bar{\mathcal{P}}_{k}\mathcal{P}^{k}
- \bar{\mathcal{P}}_{i}( n^{j} \partial_{j}C^{i} + n^{i}\partial_{j}C^{j})
+ n^{i}\bar{\mathcal{P}} \partial_{i}C
+ \mu p + \bar{\mathcal{P}}\mathcal{P}
- \mu\partial_{k}(C^{k}\bar{\mathcal{P}})
\nonumber
\\
&&
+ \sigma_{1}\pi_{i}\mathfrak{D}^{ij}\pi_{j}
+ \pi_{i}\Gamma_{1}^{i}
+ \bar{C}_{i}\{\Gamma_{1}^{i},\mathcal{H}_{k}\}C^{k}
+ \sigma_{2} Q \Delta^d Q
+ Q\Gamma_{2}
+ \bar{C}\{\Gamma_{2},\mathcal{H}_{k}\}C^{k}
\nonumber
\\
&&
+ \bar{C}_{k} \{\Gamma^{k}_{1}, p \} C
+ \bar{C}\{\Gamma_{2}, p \}C \,.
\label{gaugehamiltonian}
\end{eqnarray}
Explicitly, the brackets indicated in this Hamiltonian at quadratic order in perturbations are 
\begin{equation}
\begin{array}{l}
\bar{C}_i \{\Gamma^i_1,\mathcal{H}_k\} C^k
= 
- \sigma_{1} \bar{C}_k \left( 2\delta_{kl} \Delta^{d} 
+ (2\bar{\kappa}-1)\Delta^{d-1} \partial_{kl} \right)C^l 
\,,
\\[1ex]
\bar{C}\{\Gamma_2,\mathcal{H}_k \} C^k
=
2 \sigma_{2} \bar{C}\Delta^{d}\partial_{k}C^{k}
\,,
\\[1ex]
\bar{C}_k\{\Gamma_1^k, p \} C
=
- 2 \sigma_{1}\bar{\kappa}\bar{C}_k\Delta^{d-1}\partial_{k}C
\,,
\\[1ex]
\bar{C}\{\Gamma_2, p \} C
=
\sigma_{2} d\bar{C}\Delta^{d}C
\,,
\end{array}
\end{equation}
where $\bar{\kappa} = 1 + \kappa$. We show the quadratic potential used in (\ref{gaugehamiltonian}) in $d=2$ spatial dimensions, using the decomposition (\ref{d2decomposition}),
\begin{equation}
\mathcal{V} = \varrho(h^T\Delta^2h^T
+n\Delta^2n
-2h^T\Delta^2n) \,.
\end{equation}


\section{BRST symmetry}
The BRST symmetry in the BFV formalism is implemented by the generator $\Omega$ in the form of transformations with Dirac brackets, 
\begin{equation}
\tilde{\varphi} = \varphi + \{ \varphi \,, \Omega \}_{\mathrm{D}} \nu \,,
\label{brstdef}
\end{equation}
where $\varphi$ represents each one of the canonical fields of the fully extended canonical phase space, and $\nu$ is the fermionic global parameter of the transformation. On the canonical fields $g_{ij}$ and $\pi^{ij}$ the result of the BRST transformation is the combination of a diffeomorphism along the vector $C^i\nu$ and a conformal transformation with argument $C\nu$,
\begin{eqnarray}
&&
\delta_\Omega g_{ij} = 
\partial_k g_{ij} C^k \nu 
+ 2 g_{k(i} \partial_{j)} C^k \nu 
+ g_{ij}C\nu
\equiv \delta^{\text{diff}}_{C^{k}\nu} g_{ij}
+\delta^{\text{conf}}_{C\nu}g_{ij}\,,
\label{deltag}
\\ &&
\delta_\Omega \pi^{ij} = 
\partial_k \pi^{ij} C^k \nu 
- 2 \pi^{k(i} \partial_k C^{j)} \nu 
+ \pi^{ij} \partial_k C^k \nu 
- \pi^{ij}C\nu
\equiv \delta^{\text{diff}}_{C^{k}\nu} \pi^{ij} 
+ \delta^{\text{conf}}_{C\nu}\pi^{ij}\,.
\end{eqnarray}
The BRST transformation on $N$ results
\begin{equation}
 \delta_\Omega N =
 - W \left\{ \theta_1 \,, \mathcal{H}_i C^i + \pi C \right\} \nu \,,
 \label{deltan}
\end{equation}
where $W=\left(\frac{\delta\theta_{1}}{\delta N}\right)^{-1}$. This is close to be again a combination of a spatial diffeomorphism and a conformal transformation on $\theta_1$ (which is conformally invariant). The missing parts are that $\mathcal{H}_i$, defined in (\ref{vinmomen}), is the generator of the spatial diffeomorphisms only on the canonical pair $(g_{ij},\pi^{ij})$. The complete generator that also acts on the pair $(N,P_N)$ is
\begin{equation}
\mathcal{H}_i + \theta_2 \partial_i N \,.
\label{completegenerator}
\end{equation}
Similarly, the complete conformal generator on the pairs $(g_{ij},\pi^{ij})$ and $(N,P_{N})$ is
\begin{eqnarray}
\pi + \frac{d}{2}N\theta_{2}
\end{eqnarray}
(see Ref.~\cite{Bellorin:2018blt}). Therefore, by adding and subtracting in Eq.~(\ref{deltan}) the required terms to form these generators, we may write the transformation of $N$ as
\begin{eqnarray}
\delta_\Omega N &=&
- W \delta^{\text{diff}}_{C^{i}\nu} \theta_1
+ W \left\{ \theta_1 \,, \theta_2 \partial_i N C^i \nu \right\}
+ \frac{d}{2} W \left\{ \theta_1 \,, \theta_2 NC \nu \right\}
\nonumber \\  
&=&
- W \delta^{\text{diff}}_{C^{i}\nu} \theta_1
+  \partial_i N C^{i} \nu
+\frac{d}{2}NC\nu
\nonumber
\\
&=& - W
\delta^{\text{diff}}_{C^{i}\nu} \theta_1
+ \delta^{\text{diff}}_{C^{i}\nu} N
+\delta^{\text{conf}}_{C\nu}N\,.
\end{eqnarray}
The BRST transformations of the ghosts $C^{i}$ and $C$ can be interpreted as spatial diffeomorphisms along the vector $C^i\nu$:
\begin{eqnarray}
&&
\delta_\Omega C^i = 
\partial_j C^i C^j \nu 
\equiv \frac{1}{2} \delta^{\text{diff}}_{C^{j}\nu} C^i \,,
\label{diffC}
\\ &&
\delta_\Omega C= C^k \nu \partial_{k} C 
 = \delta^{\text{diff}}_{C^{j}\nu} C \,.
\end{eqnarray}
This diffeomorphism of $C^i$ along itself is not zero due the Grassmann nature of $C^i$ and $\nu$. The transformations of their conjugate momenta are given by
\begin{eqnarray}
\delta_\Omega \bar{\mathcal{P}}_i 
&=&
\left(
\partial_j \bar{\mathcal{P}}_i C^j
+ \bar{\mathcal{P}}_j \partial_i C^j
+ \bar{\mathcal{P}}_i \partial_j C^j
\right) \nu
+ \mathcal{H}_i \nu 
+ \bar{\mathcal{P}}\partial_{i}C \nu
\nonumber\\
& = & 
\delta^{\text{diff}}_{C^{i}\nu} \bar{\mathcal{P}}_i
+ \mathcal{H}_i \nu
+ \bar{\mathcal{P}}\partial_{i}C \nu\,,
\label{deltaP}
\\ 
\delta_\Omega \bar{\mathcal{P}} &=& 
\pi\nu
+ \left(C^k \nu \partial_k\bar{\mathcal{P}}
+ \partial_kC^k \nu \bar{\mathcal{P}}\right)
=
\delta^{\text{diff}}_{C^{k}\nu} \bar{\mathcal{P}}
+ \pi\nu \,.
\end{eqnarray} 
In summary, the BRST transformations of the fields in the extended phase space are 
\begin{equation}
\begin{array}{ll}
\delta_\Omega g_{ij} =
\delta^{\text{diff}}_{C^{k}\nu} g_{ij} 
+ \delta^{\text{conf}}_{C\nu}g_{ij}\,,
&
\delta_\Omega \pi^{ij} =
\delta^{\text{diff}}_{C^{k}\nu} \pi^{ij}
+ \delta^{\text{conf}}_{C\nu}\pi^{ij}\,,
\\[1ex]
{\displaystyle
	\delta_\Omega N =
	- W	\delta^{\text{diff}}_{C^{i}\nu} \theta_1
	+ \delta^{\text{diff}}_{C^{i}\nu} N
	+\delta^{\text{conf}}_{C\nu}N \,,
}
\quad
&
\delta_\Omega P_N = 0 \,,
\\[2ex]
\delta_\Omega N^{k} = \mathcal{P}^{k} \nu \,,
&
\delta_\Omega \pi_{k} = 0 \,,
\\[1ex] 

\delta_\Omega \mu = \mathcal{P}\nu\,,
&
\delta_\Omega Q = 0 \,,
\\[1ex]
{\displaystyle
	\delta_\Omega C^{i} = 
	\frac{1}{2} \delta^{\text{diff}}_{C^{j}\nu} C^i} \,,
& {\displaystyle
	\delta_\Omega\bar{\mathcal{P}}_{i} =
	\delta^{\text{diff}}_{C^{i}\nu} \bar{\mathcal{P}}_i
	+ \mathcal{H}_i \nu
	+ \bar{\mathcal{P}}\partial_{i}C \nu } \,, 
\\[1ex]
\delta_\Omega \mathcal{P}^{k} = 0 \,,
&
\delta_\Omega \bar{C}_{k} = 
\pi_{k} \nu \,.\\[1ex]
\delta_\Omega C = \delta^{\text{diff}}_{C^{j}\nu} C\,,
& \delta_\Omega\bar{\mathcal{P}}=\pi\nu 
+ \delta^{\text{diff}}_{C^{k}\nu} \bar{\mathcal{P}}\,, \\[1ex]
\delta_\Omega \mathcal{P} = 0 \,,
& \delta_\Omega\bar{C}=Q\nu\,. \\[1ex]
\end{array}
\label{brst}
\end{equation}

\section{Quantum Lagrangian}
Our aim in this section is to study the quantum Lagrangian, which we obtain by integration on all canonical momenta. In particular we focus on the nonlocalities and the propagators of the quantum fields. In this aspect the dimensionality of the space is very relevant, due to the conformal symmetry. We see below that we may obtain a complete set of propagators for all quantum fields in $d=2$, taking the conformal potential given in Eqs.~(\ref{potential}) and (\ref{potentiald2}). For $d \geq 3$, some fields lack their contribution to the quadratic action, due to the high order of the conformal potential and its derivatives. Hence, we cannot define propagators for these fields in perturbative theory. Motivated by this, we develop the $d=2$ case. The $d=2$ conformal theory has no local physical degrees of freedom, as general relativity. Hence, although propagators and vertices can be defined for all quantum fields, one expects that there are cancellations among the interactions of all these modes. We present the propagators of the $d=2$ case as a formal evidence of the consistency of the quantization of the theory with conformal symmetry. In the appendix we present a study of a $d=3$ conformal model deformed by nonconformal terms of lower order (that is, with explicit soft breaking of the conformal symmetry). In that model, a complete set of propagators can be obtained.

First, we can do the integration on the BFV ghost momenta $\mathcal{P}^i ,\bar{\mathcal{P}}_i , \mathcal{P}, \bar{\mathcal{P}}$ by completing the bilinear terms where they arise in the canonical action (\ref{principalaction}) and (\ref{gaugehamiltonian}). The integration on them is Gaussian and produces terms that are bilinear on the time derivatives $\dot{C}^i,\dot{\bar{C}}_i,\dot{C},\dot{\bar{C}}$. Next, we integrate in $\pi_k$ and $Q$, which leads to nonlocalities on the quantum Lagrangian,
\begin{eqnarray}
&&
\int\mathcal{D}\pi_k
\exp\left[i\int dt\,d^2x 
    \Big(-\sigma_{1}\pi_i\mathfrak{D}^{ij} \pi_j
    + \pi_k(\dot{n}^{k}-\Gamma_1^k)\Big)\right]
    =
\nonumber \\ && \qquad\qquad\qquad
    \exp\left[i\int dt\,d^2x 
    \frac{1}{4\sigma_{1}}\left(\dot{n}^i - \Gamma_1^i\right)\mathfrak{D}_{ij}^{-1}\left(\dot{n}^j - \Gamma_1^j\right)\right]
    \,,
\\ &&
\int\mathcal{D}Q
\exp\left[i\int\,dt\,d^2x 
    \left(-\sigma_{2}Q \Delta^2 Q
    + Q(\dot{\mu} - \Gamma_2)\right)\right] 
    =
\nonumber \\ && \qquad\qquad\qquad   
    \exp\left[i\int\,dt\,d^2x 
    \frac{1}{4\sigma_{2}}\left(\dot{\mu} - \Gamma_2\right)
    ( \Delta^2 )^{-1} (\dot{\mu} - \Gamma_2)\right] \,,
\end{eqnarray}
where
\begin{equation}
\mathfrak{D}^{-1}_{ij} =
\frac{\delta_{ij}}{\Delta}
- \frac{\kappa}{\bar{\kappa}}\frac{\partial_{ij}}{\Delta^2} \,.
\end{equation}
The last integration is on $p_{ij}$. To do this we use the $d=2$ decomposition
\begin{equation}
h_{ij}=
\left( \delta_{ij} - \frac{\partial_{ij}}{\Delta} \right) h^{T} 
+ \partial_{(i}h_{j)} \,,
\qquad  p_{ij}=
\left( \delta_{ij} - \frac{\partial_{ij}}{\Delta} \right) p^{T} 
+ \partial_{(i}p_{j)}\,.
\label{d2decomposition}
\end{equation}
In particular, the integration on the vector $p_i$ leads to more nonlocalities in the quantum Lagrangian. Indeed, the quadratic terms on $p_i$ arising in (\ref{principalaction}) and (\ref{gaugehamiltonian}) can be written as
\begin{eqnarray}
\frac{1}{2}\mathfrak{D}^{ij}_{0}\left(p^{i} + \mathfrak{D}_{0ik}^{-1}B_{k}\right)\left(p^{j} + \mathfrak{D}_{0jl}^{-1}B_{l}\right)
- \frac{1}{2}B_{i}\mathfrak{D}_{0ij}^{-1}B_{j} \,,
\end{eqnarray}
where
\begin{eqnarray}
&&
B_{k} = \mathfrak{D}_{0}^{kj}\left(-\frac{\dot{h}_{j}}{2} + n^{j}\right)
+ \partial_{k}\mu \,,
\\ &&
\mathfrak{D}_0^{ij} = 
\delta_{ij}\Delta + \partial_{ij} \,,
\qquad 
\mathfrak{D}^{-1}_{0ij} =
\frac{\delta_{ij}}{\Delta}
- \frac{1}{2}\frac{\partial_{ij}}{\Delta^{2}} \,.
\end{eqnarray}

We consider the measure of the second-class constraints $\delta(\theta_{1})\delta(\theta_{2}) \sqrt{\det\mathbb{M}}$ included in (\ref{measureoriginal}). The matrix of Poisson brackets of the second-class constraints has a triangular form,
\begin{equation}\label{matrixM}
\mathbb{M}=\begin{pmatrix}
\lbrace\theta_{1},\theta_{1}\rbrace & \lbrace\theta_{1},\theta_{2}\rbrace \\
-\lbrace\theta_{1},\theta_{2}\rbrace & 0
\end{pmatrix} \,.
\end{equation}
Hence the square-root factor simplifies to
\begin{equation}
\sqrt{\det\{\theta_{p}\,,\theta_{q}\}} 
=
\det \{\theta_1 \,, \theta_2 \}
\,.
\label{measuredef}
\end{equation}
This factor can be incorporated to the action by means of fermionic ghost fields $\eta\bar{\eta}$,
\begin{equation}
\det \{ \theta_1 \,, \theta_2 \} = 
\int \mathcal{D}\bar{\eta} \mathcal{D}\eta
\exp \left( i \int dt d^dx\,  
\bar{\eta} \{ \theta_1 \,, \theta_2 \} \eta 
\right) \,.
\label{bracket}
\end{equation}
The delta $\delta(\theta_1)$ can also be incorporated by means of a Lagrange multiplier $\mathcal{A}$,
\begin{equation}
\delta(\theta_1) =
\int \mathcal{D} \mathcal{A}
\exp \left( i \int dt d^dx\, \mathcal{A} \theta_1 \right)\,,
\end{equation}
similarly to the classical theory in (\ref{Hamiltonianclassical}). Finally, the quantum field $P_N$ can be eliminated by integration due to the delta $\delta(\theta_2)$ (the result of the bracket in (\ref{bracket}) is independent of $P_N$). 

After performing the above steps, we obtain the path integral in Lagrangian variables. To present it at quadratic order, we use the vectorial decomposition $h_{i} = h_{i}^{T} + \partial_{i} \Delta^{-1} h^{L}$ and $n^{i} = n^{iT} + \partial_{i} n^{L}$, with $\partial_{i}h_{i}^{T}=\partial_{i}n^{iT}=0$, and similarly for the ghost vector fields. The resulting path integral with quadratic action is
\begin{eqnarray}
Z &=& 
\int \mathcal{D}V
\exp \left\{ i \int dt d^{2}x 
\left[
- \frac{1}{4}h^{T}\left( \partial_{t}^{2}
- \left(\sigma_{2}-4\varrho\right)\Delta^{2}\right) h^{T}
+ \frac{1}{8}h_{i}^{T} \left( \partial_{t}^{2}\Delta +   
    2\sigma_{1}\Delta^{3}\right)h_{i}^{T}
\right.\right.
\nonumber \\ &&
- \frac{1}{4}h^{L}\left(\partial_{t}^{2} - \left(\sigma_{2}-4\sigma_{1}\bar{\kappa}\right)\Delta^{2}\right)h^{L}
+ \frac{\sigma_{2}}{2}h^{T}\Delta^{2}h^{L}
- \frac{1}{4\sigma_{2}}\mu\left(\partial_{t}^{2}\Delta^{-2} - 2\sigma_{2}\right)\mu
+ \mu\Delta n^{L}
\nonumber\\
&&
- \frac{1}{4\sigma_{1}}n^{iT}\left(\partial_{t}^{2}\Delta^{-1} + 2\sigma_{1}\Delta\right)n^{iT}
+ \frac{1}{4\sigma_{1}\bar{\kappa}}n^{L}\left(\partial_{t}^{2} + 4\sigma_{1}\bar{\kappa}\Delta^{2}\right)n^{L}
-\varrho n\Delta^2n
+2\varrho h^T\Delta^2n
\nonumber \\ &&
-2\varrho\mathcal{A}\Delta^2h^T
+2\varrho\mathcal{A}\Delta^2n+2\varrho\bar{\eta}\Delta^2\eta
+ \bar{C}_{k}^{T}\left(\partial_{t}^{2} + 2\sigma_{1}\Delta^{2}\right)C^{kT}
+ \bar{C}\left(\partial_{t}^{2} - 2\sigma_{2}\Delta^{2}\right)C
\nonumber \\ &&
\left.\left.
- \bar{C}^{L} \left(\partial_{t}^{2}\Delta+ 4\sigma_{1}\bar{\kappa} \Delta^{3}\right)C^{L}
- 2\sigma_{2}\bar{C}\Delta^{3}C^{L}
- 2\sigma_{1}\bar{\kappa}\bar{C}^{L}\Delta^{2}C 
\right] \right\} \,,
\label{quantumlagrangian}
\end{eqnarray}
where
\begin{equation}
 \mathcal{D}V \equiv 
 \mathcal{D}h^{T}
 \mathcal{D}h_{i}^{T}
 \mathcal{D}h^{L}
 \mathcal{D}n
 \mathcal{D}n^{kT}
 \mathcal{D}n^{L}
 \mathcal{D}\mu
 \mathcal{D}C^{iT}
 \mathcal{D}\bar{C}_{j}^{T}
 \mathcal{D}C^{L}
 \mathcal{D}\bar{C}^{L}
 \mathcal{D}C 
 \mathcal{D}\bar{C}
 \mathcal{D}\mathcal{A}
 \mathcal{D}\bar{\eta}
 \mathcal{D}\eta \,.
\end{equation}

From the above quantum Lagrangian we obtain the propagators of all fields. We present the propagators after performing a Wick rotation,\footnote{The propagators with indices are proportional to the propagators of the same objects with contracted indices. We show the contracted form in (\ref{propagators}) to reduce the size of these expressions.}
\begin{equation}
\begin{array}{rcl}
&&
\langle h^T h^T\rangle = \langle h^T n\rangle = \langle n n\rangle =
- 2\left(\omega^{2} +(4\bar{\kappa}\sigma_1-\sigma_2)k^{4}\right) \mathcal{S} \,,
\quad
\langle h^T h^L\rangle = \langle h^L n\rangle =
- 2\sigma_2 k^{4} \mathcal{S}\,,
\\[1ex] &&
k^2 \langle h_i^Th_i^T\rangle =
- 8 \mathcal{T}_3\,,
\quad
\langle h^L h^L\rangle =
- 2 \left(\omega^{2} - \sigma_{2}k^{4}\right) \mathcal{S}\,,
\quad
 \langle n^L n^L\rangle =
 2 \sigma_{1}\bar{\kappa} \left( \omega^{2} - 
2\sigma_{2}k^{4}\right) \mathcal{S} \,,
\\[1ex] &&
\langle n_i^Tn_i^T \rangle =
 4\sigma_{1}k^{2}\mathcal{T}_3\,,
\quad
\langle n^L \mu \rangle =
- 4 \sigma_{1}\sigma_{2}\bar{\kappa} k^{6} \mathcal{S} \,,
\quad
\langle \mu \mu \rangle =
- 2 \sigma_{2}k^{4} \left(\omega^{2} +
4 \sigma_{1}\bar{\kappa} k^{4}\right) \mathcal{S} \,,
\\[1ex] &&
\langle \bar{C}_{k}^{T}C^{kT}\rangle =
- 2\mathcal{T}_{3} \,,
\quad
k^2 \langle \bar{C}^{L} C^{L}\rangle =
- \left(\omega^{2} - 2\sigma_{2}k^{4}\right) \mathcal{S} \,,
\quad
\langle \bar{C}^{L} C\rangle =
2\sigma_{2} k^{4} \mathcal{S} \,,
\\[1ex] &&
\langle \bar{C}C^{L} \rangle =
- 2 \bar{\kappa}\sigma_{1} k^{2} \mathcal{S} \,,
\quad
\langle \bar{C}C\rangle =
- (\omega^2 + 4 \bar{\kappa}\sigma_{1} k^{4}) \mathcal{S}\,,
\end{array}
\label{propagators}
\end{equation}
and 
\begin{equation}
 \langle\mathcal{A}\mathcal{A}\rangle =
 \langle n\mathcal{A} \rangle =
 \langle\eta\bar{\eta} \rangle =
  \frac{1}{\varrho k^{4}}\,, 
 \label{irregular}
\end{equation}
where
\begin{eqnarray}
&&
\mathcal{T}_1 = 
\left[\omega^2 
+ \left(2\sigma_{1}\bar{\kappa} - \sigma_{2} + \sqrt{4\bar{\kappa}^{2}\sigma_{1}^{2}+\sigma_{2}^{2}}\right) k^{4}\right]^{-1} \,,
\label{T1}
\\ &&
\mathcal{T}_2 = 
\left[\omega^2 
+ \left(2\sigma_{1}\bar{\kappa} - \sigma_{2} - \sqrt{4\bar{\kappa}^{2}\sigma_{1}^{2}+\sigma_{2}^{2}}\right)
k^{4}\right]^{-1} \,,
\label{T2}
\\ &&
\mathcal{T}_3 = \left(\omega^2 + 2\sigma_{1}k^4\right)^{-1}\,,
\\ &&
\mathcal{S} \equiv 2 \mathcal{T}_1 \mathcal{T}_2 \,.
\label{T3}
\end{eqnarray}
The propagators $\mathcal{T}_1$, $\mathcal{T}_2$, and $\mathcal{T}_3$ can be made regular (with positive coefficients) \cite{Anselmi:2008bq} if the coupling constants satisfy
\begin{equation}
 \bar{\kappa} > 0 \,,
 \qquad
 \sigma_1 > 0 \,,
 \qquad
 \sigma_2 < 0 \,.
\end{equation}
$\mathcal{T}_1$, $\mathcal{T}_2$, and $\mathcal{T}_3$ determine the propagators of the fields in (\ref{propagators}); hence all propagators in (\ref{propagators}) are regular if this condition is satisfied. The auxiliary fields $\mathcal{A}$ and $\eta\bar{\eta}$ acquire irregular propagators. These fields are associate to the measure of the second-class constraints, and the irregularity of their propagators was previously observed in the general nonprojectable theory without the conformal symmetry.

\section{Conclusions}
We have seen that the gauge symmetry of anisotropic Weyl transformations can be incorporated to the BFV quantization of the nonprojectable Ho\v{r}ava theory. This provides a quantum theory with conformal transformations different to the relativistic case. We have introduced a local gauge-fixing condition for the symmetry of spatial diffeomorphisms and the Weyl transformations. The way we introduce the gauge fixing is inspired by the nonlocal gauge fixing-condition of the projectable case \cite{Barvinsky:2015kil}, which is known to have a local counterpart in the BFV formalism \cite{Bellorin:2021udn}. The gauge-fixing condition naturally splits the metric variables between the vector longitudinal sector for the spatial diffeomorphisms and the trace for the Weyl transformations. On top of this, we have seen that the quantization with the two gauge symmetries is compatible with the second-class constraints in the BFV formalism. We have expressed the BRST transformations of the fields mainly in terms of a spatial diffeomorphisms and an anisotropic Weyl transformation, with the ghost fields entering in the arguments of the transformations.

A feature of the quantization we have studied is the Feynman rules. We have obtained the quantum Lagrangian by integration on the BFV canonical path integral in the $2+1$ case. We have found some features similar to the case of the nonconformal theory. The quantum Lagrangian gets nonlocalities as a consequence of the gauge-fixing procedure. Most of the fields gets regular propagators \cite{Anselmi:2008bq}, a feature relevant for the renormalizability of the theory \cite{Barvinsky:2015kil}. The only exception are the fields associated to the measure of the second-class constraints. This behavior is qualitatively identical to the general nonprojectable theory without the conformal symmetry. We have previously found that the divergences coming from the irregular loops cancel each other exactly in the nonprojectable theory \cite{Bellorin:2021udn}. Therefore, a natural question for a future work is whether cancellations of this kind can also occur in the conformal theory. In the $3+1$ case or higher the definition of Feynman rules is more difficult since the conformal potentials do not contribute to the quadratic action, hence some propagators of fields are lacked (the propagators of the fields associated with the measure of the second-class constraints, $\mathcal{A}$ and $\eta,\bar{\eta}$, cannot be defined since these fields decouple from the quadratic action due to the order of the conformal potential in $3+1$ and higher).


\section*{Acknowledgments}
C.B.~is partially supported by Grant No.~CONICYT PFCHA/DOCTORADO BECAS CHILE /2019 -- 21190960. C.B.~is partially supported as graduate student in the ``Doctorado en F\'isica Menci\'on F\'isica-Matem\'atica" Ph.D. program at the Universidad de Antofagasta. 

\appendix

\section{Breaking the conformal symmetry}
The potential $\mathcal{V} = \varrho\chi^d$ in (\ref{potential}) contributes at quadratic order in perturbations only for the spatial dimension $d=2$. A technical issue with this is that the Lagrange multiplier $\mathcal{A}$ contributes to the quadratic action only for the $d=2$ dimensionality; hence only in this case one may define propagators for the $\mathcal{A}$ field. This is a consequence of the fact that $\mathcal{A}$ multiplies the constraint $\theta_1$, which is a derivative of the primary Hamiltonian, and hence of $\mathcal{V}$. Therefore, the conformal model we have presented in this paper, although it is consistent in the $d=2$ case as we have shown, is not totally suitable for doing Feynman diagrams in $d \geq 3$ dimensionality, whenever one wants to use the $\mathcal{A}$ field. On the other hand, in $d=3$ the conformal potential $(\text{Cotton-tensor})^2$ does contribute to the quadratic action, but this does not cure the problem of the decoupling of $\mathcal{A}$ at the quadratic level. The reason in this case is that the $(\text{Cotton-tensor})^2$ potential does not depend on $N$ at quadratic level; hence $\theta_1$ does not get linear-order contributions. Therefore, to have a well-posed perturbation theory in $d=3$, we study explicit deformations of the conformal symmetry.

The deformation we make on the conformal theory consists of adding nonconformal terms of order $z=1$, and moving the coupling constant $\lambda$ from its critical value, $\lambda \neq 1/d$. At the end of the analysis we discuss on the limit when $\lambda$ goes back to its critical value. The constraint $\pi = 0$ does not arise in this modified theory, hence the theory does not possess a conformal-symmetry generator. 

For concreteness, we specialize to the $d=3$ case and with the conformal potential made with the Cotton tensor \cite{Horava:2009uw}, which is given by
\begin{eqnarray}
C^{ij}=\frac{1}{\sqrt{g}}\varepsilon^{kl(i}\nabla_{k} R^{j)}{}_l \,.
\end{eqnarray}
It transforms under Weyl transformation as $\tilde{C}^{ij}=\Theta^{-5}C^{ij}$. The conformal potential
\begin{equation}
 \mathcal{V} = C^{ij}C_{ij} \,,
\end{equation}
whose conformal weight is $-3$, can be introduced in the primary Hamiltonian. The deformation consists of including in the potential the known $z=1$ terms of the nonprojectable Ho\v{r}ava theory. Thus, we take the primary Hamiltonian density as
\begin{equation}
 \mathcal{H}_0 = 
 \frac{N}{\sqrt{g}}\pi^{ij}\pi_{ij}
 + \zeta \sqrt{g} N C^{ij}C_{ij}
 - \sqrt{g}N ( \beta R + \alpha a_i a^i )
 \,.
\end{equation}
The deformed theory has the following constraints:
\begin{eqnarray}
\mathcal{H}_i &=& - 2 g_{ij} \nabla_k \pi^{jk} \,,
\\
\theta_1 &=& 
\frac{N}{\sqrt{g}}\pi^{ij}\pi_{ij}
+ \zeta \sqrt{g} N C^{ij}C_{ij}
- \sqrt{g}N(\beta R+\alpha a_i a^i )
-2\alpha\sqrt{g}\Delta N \,,
\\
\theta_2 &=& P_N \,.
\end{eqnarray}

We perform the BFV extension of the phase space. The additional canonical pairs are $(N^{i},\pi_{i})$ and $\eta^{a}=( C^i,\mathcal{P}^i)$, $\mathcal{P}_{a}=(\bar{\mathcal{P}}_{i}, \bar{C}_i)$ The BRST charge is given by
\begin{eqnarray}
\Omega &=&\int\,d^{3}x\left(
\mathcal{H}_{k}C^{k}
+ \pi_{k}\mathcal{P}^{k}
- C^k\partial_kC^l\bar{\mathcal{P}}_l\right)
\,.
\end{eqnarray}
For the gauge-fixing fermionic function we use the form $\Psi = \bar{\mathcal{P}}_i n^i
+ \bar{C}_i \chi_1^i$. Since this is the gauge-fixing corresponding to the symmetry of spatial diffeomorphisms, we use the same factor $\chi_1^i$ given in (\ref{chi1}). In this model we set the constants
$c_{1}=-\sigma_{1}$, $c_{2}=\lambda\bar{\kappa}\sigma_{1}$, $c_{3}=-2\kappa\sigma_{1}$ to cancel odd derivatives in the Lagrangian. Thus, the gauge-fixed Hamiltonian takes the form
\begin{equation}
{\mathcal{H}}_{\Psi} 
=
\mathcal{H}_{0}
+ \mathcal{H}_{k}n^{k}
+ \bar{\mathcal{P}}_{k}\mathcal{P}^{k}
- \bar{\mathcal{P}}_{i} (n^{j}\partial_{j}C^{i} + n^{i}\partial_{j}C^{j})
+ \sigma_1\pi_{i}\mathfrak{D}^{ij}\pi_{j} 
+ \pi_{i}\Gamma_{1}^{i}
+ \bar{C}_{i}\{\Gamma_{1}^{i},\mathcal{H}_{k}\}C^{k} \,.
\end{equation}

To integrate all the canonical momenta we use the following decomposition of tensors in $d=3$,
\begin{equation}
h_{ij}=	
h_{ij}^{TT} 
+ \frac{1}{2}\Big( \delta_{ij} 
- \frac{\partial_{ij}}{\partial^2}\Big)h^T
+ \partial_{(i}h_{j)}
\,,\,\,\,\text{with}\,\,\,\,
h_{ii}^{TT}=\partial_{i}h_{ij}^{TT}=0 \,.
\end{equation}

To obtain the propagators we make the following decomposition on vectors:
$h_{i}=h_{i}^{T}+ \Delta^{-1} \partial_{i}h^{L}$ and $n^{i}=n^{iT}+\partial_{i}n^{L}$, with $\partial_{i}h_{i}^{T}=\partial_{i}n^{iT}=0$. 
 The decomposed path integral is
\begin{eqnarray}
 		Z &=& 
 		\int
 		\mathcal{D}h_{ij}^{TT}
 		\mathcal{D}h^{T}
	    \mathcal{D}h_{i}^{T}
	    \mathcal{D}h^{L}
	    \mathcal{D}n
 		\mathcal{D}n^{kT}
 		\mathcal{D}n^{L}
        \mathcal{D}C^{iT}
 		\mathcal{D}\bar{C}_{j}^{T}
        \mathcal{D}C^{L}
        \mathcal{D}\bar{C}^{L}
 		\mathcal{D}\mathcal{A}
 		\mathcal{D}\bar{\eta}
 		\mathcal{D}\eta
 		\nonumber
 		\\ 
 		&&
 		\exp\left\lbrace i \int dt d^{3}x 
 		\left[\frac{1}{4}h_{ij}^{TT}\left(-\partial_{t}^{2}
 		+ \beta\Delta
 		+ \zeta\Delta^{3}\right)h_{ij}^{TT}
 		- \frac{1}{8}h^{T}\left[(1-2\lambda)\partial_{t}^{2}
 		+ \beta \Delta
        + 8\lambda^{2}\bar{\kappa}\sigma_{1}\Delta^{3}\right]h^{T}
 		\right.\right.
 		\nonumber\\
 		&&
 		+ \frac{1}{8}h_{i}^{T}\left(\partial_{t}^{2}\Delta + 2\sigma_1\Delta^{4}\right)h_{i}^{T}
 		- \frac{(1-\lambda)}{4}h^{L}\left[\partial_{t}^{2} + 4\sigma_1\bar{\kappa}(1-\lambda)\Delta^{3}\right]h^{L}
        \nonumber\\
        &&
        + \frac{\lambda}{2}h^{T}\left[\partial_{t}^{2} + 4\sigma_1\bar{\kappa}(1-\lambda)\Delta^{3}\right]h^{L}
        - \alpha n\Delta n
        - \beta n\Delta h^{T}
 		\nonumber\\
 		&&
 		- \frac{1}{4\sigma_{1}}n^{iT}\left(\partial_{t}^{2}\Delta^{-2} + 2\sigma_1\Delta\right)n^{iT}
 		+ \frac{1}{4\sigma_{1}\bar{\kappa}}n^{L}\left[\partial_{t}^{2}\Delta^{-1}
        + 4\bar{\kappa}\sigma_1(1-\lambda)\Delta^{2}\right]n^{L}
	    \nonumber
	    \\
	    &&
	    + \bar{C}_{k}^{T}\left(\partial_{t}^{2}+2\sigma_1\Delta^{3}\right)C^{kT}
	    - \bar{C}^{L}\left[\partial_{t}^{2}\Delta + 4\sigma_1\bar{\kappa}(1-\lambda)\Delta^{4}\right]C^{L}
	    \nonumber
	    \\
	    && \left.
 		+ \mathcal{A}\left(\beta\Delta h^{T}+2\alpha\Delta n\right)
 		- 2\alpha\bar{\eta}\Delta\eta
 		\Big]\right\}
 		\,.
 \end{eqnarray}

The propagators of the deformed theory, after a Wick rotation, are given by
\begin{equation}
\begin{array}{rcl}
&&
\langle h_{ij}^{TT}h_{ij}^{TT}\rangle =
- 4\mathcal{T}_{1} \,,
\quad
\langle h^T h^T\rangle = 
- 8(1-\lambda) \mathcal{T}_2\,,
\quad
k^2 \langle h_i^Th_i^T\rangle =
- 8 \mathcal{T}_3\,,
\\[1ex] &&
\langle h^L h^L\rangle =
4 \left[(2\lambda-1)\omega^{2} + \beta(1-\frac{2\beta}{\alpha})k^{2} + 8\sigma_{1}\bar{\kappa}\lambda^{2}k^{6}\right]\mathcal{T}_2\mathcal{T}_4\,,
\quad
\langle h^T h^{L}\rangle =
- 8\lambda \mathcal{T}_2\,,
\\[1ex] &&
\langle h^L n\rangle =
\frac{4\lambda\beta}{\alpha}\mathcal{T}_2\,,
\quad
\langle h^T n\rangle =
\frac{4(1-\lambda)\beta}{\alpha}\mathcal{T}_2\,,
\quad
\langle n n\rangle =
- \frac{2(1-\lambda)\beta^2}{\alpha^{2}}\mathcal{T}_2\,,
\\[1ex] &&
\langle n_i^Tn_i^T \rangle = 
- 4\sigma_{1}k^4\mathcal{T}_3\,,
\quad
\langle n^L n^L\rangle =
- 4 \bar{\kappa}\sigma_1 k^2 \mathcal{T}_4 \,,
\\[1ex] &&
\langle \bar{C}_{k}^{T}C^{kT}\rangle =
- 2 \mathcal{T}_{3} \,,
\quad
k^2 \langle \bar{C}^{L} C^{L}\rangle =
- 2 \mathcal{T}_{4} \,,
\end{array}
\label{propagatorsdeformed}
\end{equation}
and
\begin{equation}
\langle\mathcal{A}\mathcal{A}\rangle =
\langle n\mathcal{A} \rangle =
\langle\bar{\eta}\eta \rangle =
- \frac{1}{\alpha k^2}\,,
\label{irregulardeformed}
\end{equation}
where
\begin{eqnarray}
\mathcal{T}_1 &=& \left(\omega^2 + \beta k^{2} +\zeta k^{6}\right)^{-1},
\\
\mathcal{T} _2 &=& 
\left[ ( 1 - 3\lambda ) \omega^2 - \beta ( 1-\lambda ) \left(1-\frac{2\beta}{\alpha}\right) k^2\right]^{-1} ,
\label{T2deformed}
\\
\mathcal{T}_3 &=& (\omega^2-2\sigma_{1}k^6)^{-1}\, ,
\\
\mathcal{T}_4 &=& (\omega^2-4\bar{\kappa}\sigma_{1}(1-\lambda)k^6)^{-1}.
\end{eqnarray}
In this deformed theory, the regularity of the propagators in (\ref{propagatorsdeformed}) is achieved with the conditions
\begin{equation}
 \beta > 0 \,,
 \quad
 \zeta > 0 \,,
 \quad
 \sigma_1 < 0 \,, 
 \quad 
 1-3\lambda > 0 \,,
 \quad
 \bar{\kappa}(1-\lambda) > 0\,,
 \quad
 (1-\lambda)\left(1-\frac{2\beta}{\alpha}\right)<0\,.
\end{equation}
We may regard as physical modes the ones corresponding to the propagators $\mathcal{T}_1$ and $\mathcal{T}_2$, since $\mathcal{T}_3$ and $\mathcal{T}_4$ enter in the propagators of the ghosts. $\mathcal{T}_1$ yields the propagators of the two tensorial modes in $h_{ij}^{TT}$ and $\mathcal{T}_2$ is the so called extra scalar mode of the Ho\v{r}ava theory (it arises it several quantum fields due to the gauge we are using).

Now we can go to the critical point $\lambda = 1/3$, where the classical theory acquires Weyl symmetry on its kinetic term. $\mathcal{T}_2$ lacks its temporal dependence in this limit. This is a reminiscence of the dynamics of the critical theory, which does not necessarily have a conformal potential. Due to the critical value of $\lambda$, there arise more second-class constraints, which is in agreement with the fact that the would be extra scalar mode lacks its kinetic term. Therefore, a physical mode in this limit is eliminated and there remains the modes of $\mathcal{T}_1$ as physical modes. This effect recalls that we are approaching the conformal symmetry; hence the elimination of the extra mode. Moreover, the $\mathcal{T}_1$ propagator admits the limit $\beta=\alpha = 0$, where the exact conformal symmetry is recovered. But the vanishing of $\alpha$ affects the definition of the propagators involving the $\mathcal{A}$ field in (\ref{irregulardeformed}). This is natural, since the $z=1$ terms of deformation were introduced to get propagators for this field, as we have discussed.


\end{document}